\documentclass[aps,twocolumn,superscriptaddress]{revtex4-1}
\usepackage{graphicx,hyperref,epstopdf,amsfonts,amsmath}
\usepackage{amssymb}
\usepackage{upgreek}
\usepackage{color}
\usepackage{epstopdf}

\begin{document}

\title{Facilitation  of  DNA  loop  formation  by  protein-DNA
non-specific interaction}

\author{Jaeoh Shin}
\affiliation{Department of Chemistry, Rice University, Houston, Texas, 77005, USA}
\author{Anatoly B. Kolomeisky} 
\affiliation{Department of Chemistry, Rice University, Houston, Texas, 77005, USA}%
\affiliation{Department of Chemical and Biomolecular Engineering, Rice University, Houston, Texas, 77005, USA}
\affiliation{Center for Theoretical Biological Physics, Rice University, Houston, Texas, 77005, USA}

\begin{abstract}
Complex DNA topological structures, including polymer loops, are frequently observed in biological processes when protein molecules simultaneously bind to several distant sites on DNA. However, the molecular mechanisms of formation of these systems remain not well understood. Existing theoretical studies focus only on specific interactions between protein and DNA molecules at target sequences. However, the electrostatic origin of primary protein-DNA interactions suggests that interactions of proteins with all DNA segments should be considered. Here we theoretically investigate the role of non-specific interactions between protein and DNA molecules on the dynamics of loop formation. Our approach is based on analyzing a discrete-state stochastic model via a method of first-passage probabilities supplemented by Monte Carlo computer simulations. It is found that depending on a protein sliding length during the non-specific binding event three different dynamic regimes of the DNA loop formation might be observed. In addition, the loop formation time might be optimized by varying the protein sliding length, the size of the DNA molecule, and the position of the specific target sequences on DNA. Our results demonstrate the importance of non-specific protein-DNA interactions in the dynamics of DNA loop formations. Several quantitative predictions that can be experimentally tested are also presented.
\end{abstract}

\maketitle

%%%END OF FOOTNOTES%%%
\section{Introduction}
Many biological phenomena involve the formation of complex topological structures, which are typically made of protein and nucleic acid biopolymers.\cite{albert02} In most cases, this is a result of proteins binding simultaneously to spatially distant specific target sites on DNA, which leads to the appearance of DNA loops. \cite{matthews1992, schleif1992} Specific biological processes with the formation of DNA loops include gene regulation and gene rearrangements via site-specific recombination. \cite{halford2004,grindley2006mechanisms,mani2010triggers,bushman2005genome,cournac2013dna} Due to its fundamental importance in natural systems, many theoretical models were proposed to describe the loop formation process in polymer systems. \cite{wilemski1974, wilemski1974b, szabo1980,toan2008, guerin2012non,saiz2006dna,allemand2006}  It also was extensively studied experimentally using various techniques.\cite{finzi1995, bonnet1998,chen2010,allemand2006,stiehl2013kinetics} In addition, many recent investigations considered the loop formation in biologically relevant settings, such as in crowded environment,\cite{stiehl2013kinetics, shin2015a} in confined medium\cite{amitai2013diffusing, shin2015b} and in the presence of non-equilibrium fluctuations.\cite{shin2012} However, many aspects of the dynamics of loop formation remain not clarified.

While the molecular mechanism of the DNA loop formation by multi-site proteins is not fully understood, it is reasonable to assume that the protein molecule that has several DNA binding sites first attaches to one of the specific sites on DNA, and subsequently it sequentially associates to the other sites. In the majority of previous theoretical studies, it is assumed that the protein interacts only with the specific target sequences on DNA.\cite{chen2014,mulligan2015} However, as the dominating interaction between the protein and DNA is of the electrostatic origin, \cite{privalov2010} it seems reasonable to suggest that the protein-DNA non-specific interactions might also be important. In this scenario, the protein already bound to DNA at one site can bind to a random site of the DNA, forming a transient loop, and the protein then diffuses (slides) along the strand searching for the target site. If the target is not found, the protein dissociates and the process is repeated until the target sequence is located. Indeed, this idea is known as a facilitated diffusion in the process of protein search for a target sequence, and it was shown to be important for single-site proteins that do not form DNA loops. The combination of three-dimensional (3D) diffusion in bulk and one-dimensional (1D) sliding can dramatically enhance the effective protein-DNA association rates. \cite{riggs1970lac,berg1976association,winter1981diffusion, von1989facilitated,coppey2004,lomholt2009,kolomeisky2013b} The facilitated diffusion in biologically systems has been studied extensively in the past several decades, and it is reviewed, for instance, in Ref.\cite{halford2004,mirny2009,benichou2011,kolomeisky2011,sheinman2012}

Recently, we theoretically investigated the role of transient DNA looping on the search dynamics for specific targets on DNA by multi-site proteins.\cite{kolomeisky2016} It was shown using analytical calculations and computer simulations that the formation of DNA loops might accelerate the overall search process. However, the role of the protein sliding in the context of polymer loop formation has not been studied so far. At the same time, experiments clearly show that proteins might translocate along the DNA chain while being in the looped conformation.\cite{gilmore2009single}

In this paper, we present a theoretical approach to investigate the protein-mediated loop formation kinetics, which also directly incorporates the sliding along the DNA chain. It is assumed that the protein molecule has two DNA-binding sites, and one of them is already bound to the end of the DNA molecule. It remains bound all the time while the search for the second target sequence is taking place. Because the protein is already bound to DNA at one site, the non-specific protein-DNA interactions depend on the loop size. Therefore, one cannot use theoretical approaches developed for the binding of the single-site protein to target sites. \cite{halford2004,mirny2009,kolomeisky2011}  To explain the dynamics of the system, we take into account the free energy cost of the loop formation. It is found that depending on the protein sliding length, which is the average length that the protein moves along DNA during one binding cycle, the loop formation process shows different dynamic behaviors. Moreover, the loop formation time can be minimized at an intermediate value of the sliding length. The specific location of the target site and the length of the DNA segment also influence the search process. Our results indicate that the non-specific protein-DNA interactions play an essential role in the polymer loop formation.

The paper is organized as follows. The theoretical model is described in the Sec. \ref{sec-model}, and analytic results in limiting cases are presented in Sec. \ref{sec-analytic}. The general results are presented and discussed in Sec. \ref{sec-results}, and we summarize and conclude in Sec. \ref{sec-conclusion}.

\section{Theoretical model}
\label{sec-model}

\begin{figure}
\centering
\includegraphics[width=0.95\columnwidth]{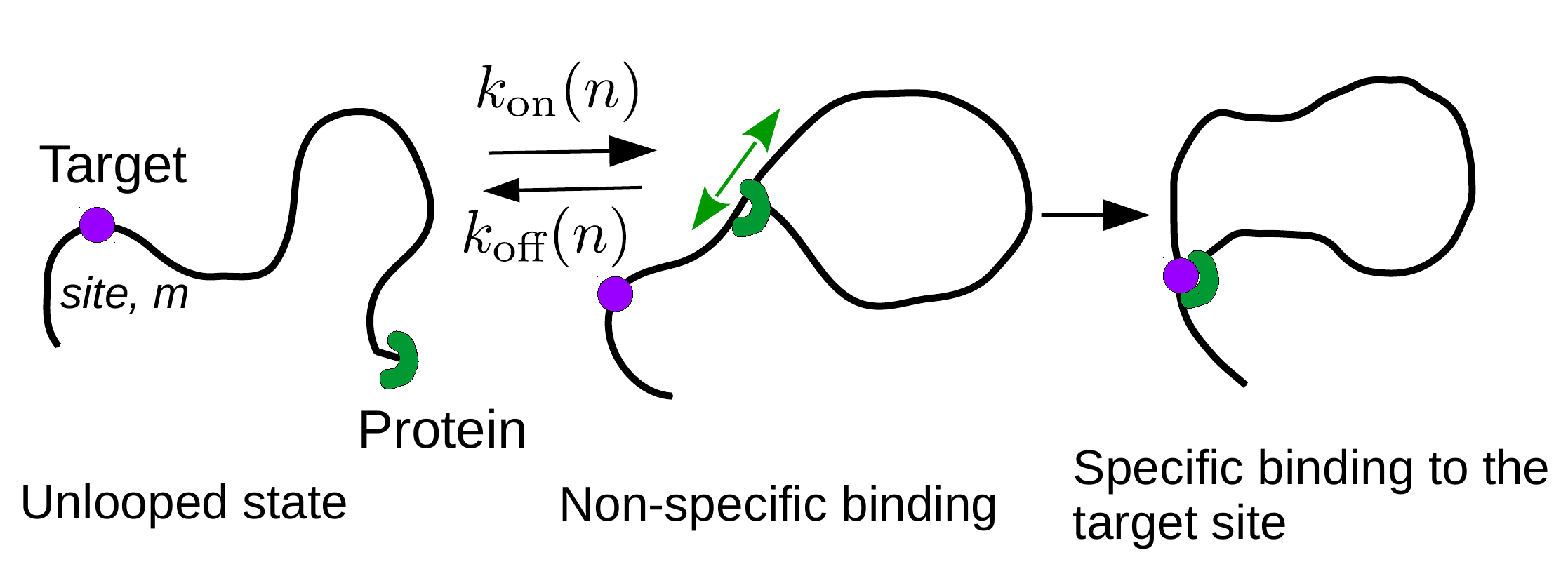}
\includegraphics[width=0.85\columnwidth]{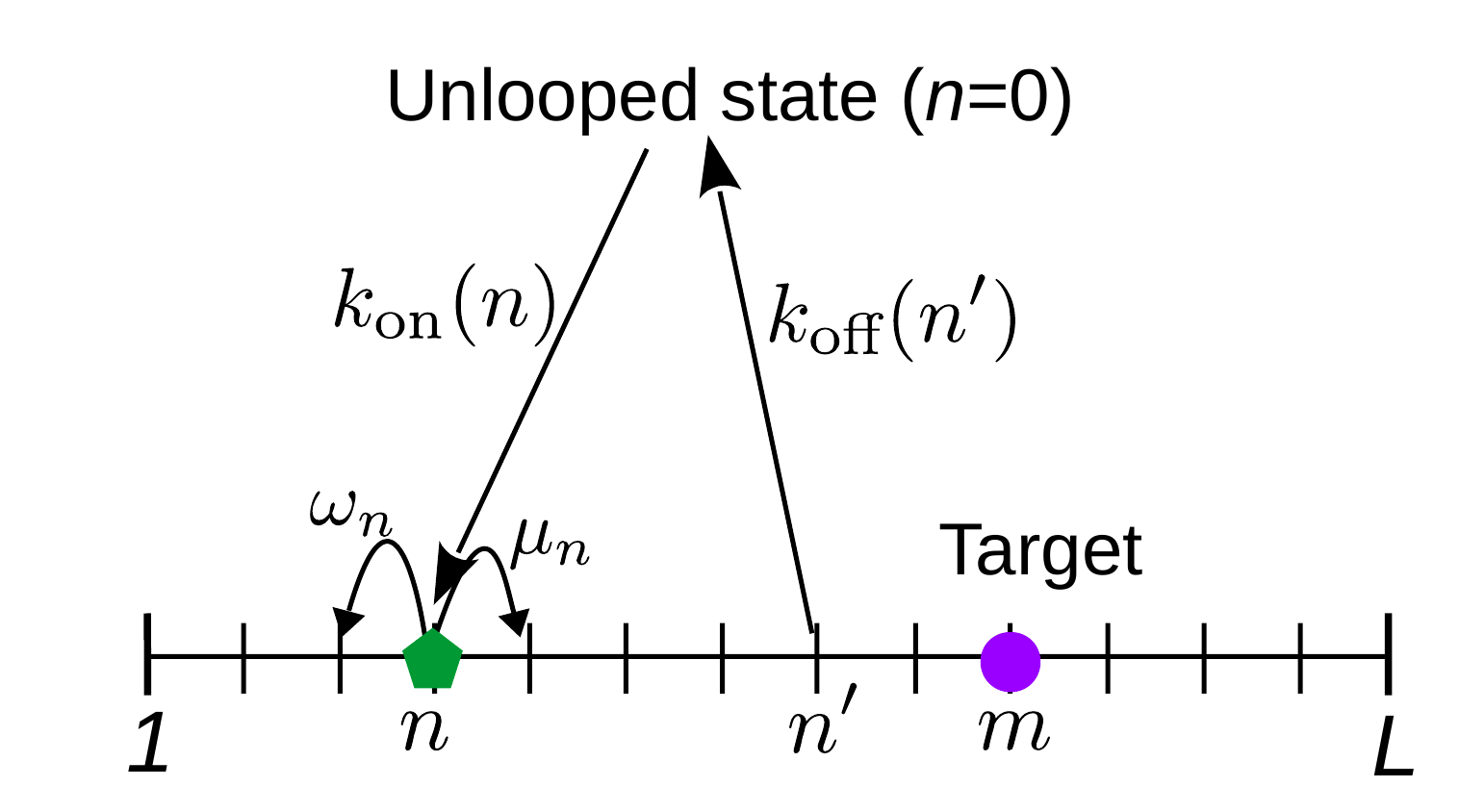}
\caption{(Top) Schematic view of the DNA looping process. Here the multi-site protein molecule (green), already bound to one end of the DNA, is searching for a target site (violet). (Bottom) The discrete-state stochastic model of the search process.}
\label{fig-1}
\end{figure}

Let us consider a process of the protein searching for a target sequence on DNA as illustrated in Fig. \ref{fig-1} Top. It is assumed here that the protein is already bound to one end of the DNA chain (and remains there forever) while exploring the space to find the second binding site on the same strand. This is a reasonable assumption because specific protein-DNA interactions are very strong.\cite{halford2004} The system can be viewed as $L+1$ discrete states, see Fig. \ref{fig-1} Bottom. If the protein is in the state $1 \le n \le L$ it means that the DNA loop of size $n$ is formed and the DNA segment of length $L-n$ is free. The final target sequence is in the state $m \neq 0$. The state $n=0$ corresponds the protein molecule unbound from the DNA chain (but still connected to the DNA end site). The protein can non-specifically associate to the state $n$ with a rate $k_{\text{on}}(n)$, while the dissociation rate is equal to  $k_{\text{off}}(n)$ (Fig. \ref{fig-1}). The non-specific binding energy (enthalpic contribution) is given by $\epsilon$ ($\epsilon<0$ corresponds to attraction and $\epsilon>0$ corresponds to repulsion). This also means that we are neglecting the effect of DNA sequence heterogeneity, although it might be relevant.\cite{shvets2015} In the non-specifically bound state, the protein can diffuse along the chain with the position-dependent rates that also depend on the direction of the motion (see Fig. \ref{fig-1}). The process of reducing the size of DNA loop is taking place with a rate $w_{n}$, while increasing the loop size is associated with a rate $\mu_{n}$.

Assuming that the relaxation of the DNA chain is taking place faster than any other processes in the system, the dynamics is governed by changes in the free energy. At realistic cellular conditions, a significant fraction of the free energy is due to the formation and breaking of DNA loops. The free energy cost of forming a loop of size $n$ (in the unit of thermal energy, $k_{\text{B}}T$) is\cite{kolomeisky2016}
\begin{equation}
G_0(n)=\frac{A}{n}+\alpha \log[n].
\end{equation}
In this expression, the first term accounts for the polymer bending energy and the second term describes the entropic cost of the loop formation. The coefficient $A$ is proportional to the bending stiffness of the DNA chain. For instance, for the case of a circular loop, $A=2\pi^2 l_p$, where $l_p$ is the persistence length of the chain in the units of base-pair length. The exponent $\alpha$ is related to the scaling exponent for the radius of gyration, and for the ideal Gaussian chain it is equal to $\alpha=3/2$. Although there are more advanced models of the polymer looping,\cite{chen2014, mulligan2015} it is expected that our simplified model still should account for the main physical features of the search process with loop formation.

\begin{figure}
\centering
\includegraphics[width=0.85 \columnwidth]{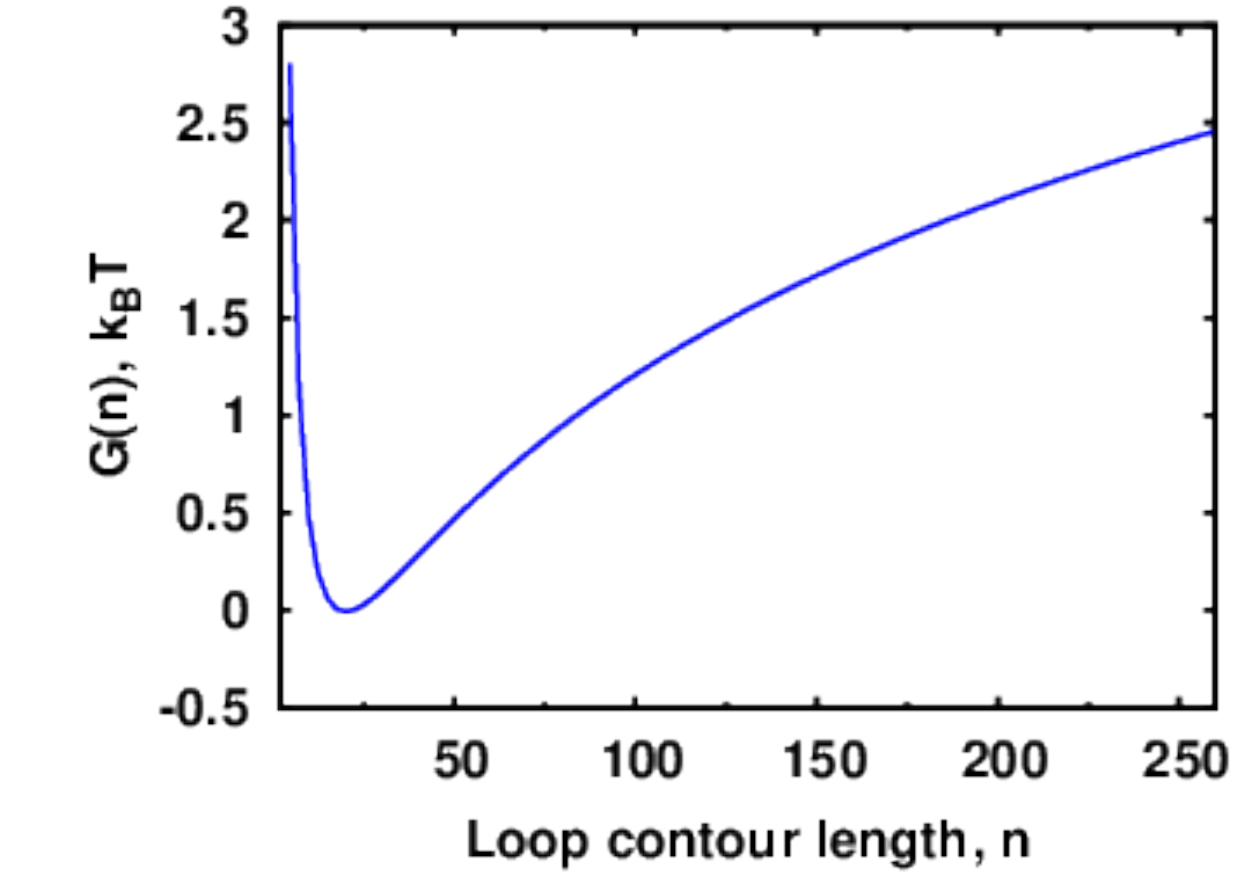}
\caption{Free-energy cost of the DNA loop formation as a function of the loop contour length $n$. In our calculations, we assume that $A=30$ and $\alpha=3/2$. }\label{fig-2}
\end{figure}

The total free energy cost of loop formation should also include the enthalpic contribution due to the protein-DNA non-specific binding energy, and the final expression is given by
\begin{equation}
G(n)=G_0(n)+\epsilon=\frac{A}{n}+\alpha \log[n]+\epsilon.
\end{equation}
The specific example of the free-energy profile is given in Fig. \ref{fig-2}. This allows us to evaluate the position-dependent binding and unbinding rates:
\begin{equation}
 k_{\text{on}}(n)=k_{\text{on}}^{(0)} \exp\left[-\theta G_0(n)\right],
\end{equation}
and
\begin{equation}
k_{\text{off}}(n)=k_{\text{off}}^{(0)}\exp\left[(1-\theta)G_0(n)\right],
\end{equation}
where $k_{\text{on}}^{(0)}$ and $k_{\text{off}}^{(0)}$ are association and dissociation rates, respectively, in the absence of loop formation. The parameter $ 0 \le \theta \le 1$ reflects the relative contribution of free energy changes to the binding and unbinding rates. It can be shown that the main results of this work do not depend much on the specific value of $\theta$, and, to simplify calculations, we set $\theta=0.5$ from now on. Detailed balance arguments suggest that binding/unbinding rates are related to each other as
\begin{equation}\label{eq5}
\frac{k_{\text{on}}^{(0)}}{k_{\text{off}}^{(0)}}=\exp(-\epsilon),
\end{equation}
which leads to
\begin{equation}\label{eq6}
\frac{k_{\text{on}}(n)}{k_{\text{off}}(n)}=\exp\left[-G(n)\right].
\end{equation}
The physical interpretation of Eqs. (\ref{eq5}) and (\ref{eq6}) is simple. If the formation of the DNA loop lowers the free energy of the system, then the corresponding association rate is faster and breaking the loop is a slower process. But if the formation of the DNA loop increases the free energy of the system, then the corresponding binding rate is slow while the unbinding transition is fast. 

The direction-dependent diffusion of protein along the DNA chain is affected by the free-energy changes associated with varying the size of DNA loops. More specifically, we can write
\begin{equation}
\mu_{n}=\mu_0\exp\left[-\theta \Delta G(n+1)\right]; \quad \omega_n=\mu_0\exp\left[(1-\theta) \Delta G(n)\right],
\end{equation}
where $\mu_n$ ($\omega_n$) is the sliding rate that makes the loop size increasing (decreasing) by one unit length, and 
\begin{equation}
\Delta G(n) \equiv G(n)-G(n-1) = G_{0}(n)-G_{0}(n-1),
\end{equation}
is the associated free-energy difference. The sliding rate $\mu_{0}$ describes the diffusion in the absence of the loop formation, i.e., in the flat free-energy profile, and it is generally determined by the value of the enthalpic energy $\epsilon$. We again assume that $\theta=1/2$ to simplify computations. In addition, the sliding rates are  related to each other via the detailed balance arguments,
\begin{equation}\label{eq8}
\frac{\mu_{n-1}}{ \omega_n}=\exp\left[-\Delta G(n)\right].
\end{equation}
This expression implies that the protein sliding is faster in the direction of lowering the free energy of the system, while the sliding is slower in the direction of increasing the free energy of the system.

To analyze the dynamics of the polymer loop formation by the multi-site protein, a method of first-passage probabilities, which have been successfully employed in studies of various protein search processes for target sites,\cite{kolomeisky2013b,kolomeisky2016, shvets2015,shin2018a,shin2018b} is utilized.  We define a first-passage time probability density function $F(n,t)$, which describes the probability to reach the target site $m$ at time $t$ given that it was at the site $n$ at $t=0$. The state  $n=0$ is the unbound state (see Fig.\ref{fig-1} A). The temporal evolution of the first-passage probabilities $F(n,t)$ follows the backward master equations,\cite{kolomeisky2013b,kolomeisky2016}
\begin{eqnarray}
    \frac{\partial F(n,t)}{\partial t}  & = & -\left[\mu_n+\omega_n+k_{\text{off}}(n)\right] F(n,t) +\mu_nF(n+1,t)+\omega_nF(n-1,t) \nonumber \\
    & & +k_{\text{off}}(n)F(0,t),
\label{eq-M5}
\end{eqnarray}
for $n\neq m$. For $n=0$ state we have,
\begin{equation}
    \frac{\partial F(0,t)}{\partial t}=-F(0,t) \sum_{n=1}^{L} k_{\text{on}}(n) + \sum_{n=1}^{L} k_{\text{on}}(n) F(n,t).
\label{eq-M6}
\end{equation}
Additionally, the initial condition implies that $F(m)=\delta(t)$, which means that if the protein is at the site $m$ at time $t=0$, the process will end immediately. Calculating explicitly these first-passage probabilities should provide a full dynamic description of the system.\cite{kolomeisky2013b,kolomeisky2016}

\section{Dynamics in limiting cases}
\label{sec-analytic}

Although we were not able to determine the first-passage probabilities explicitly in general situations, there are several limiting cases that can be solved analytically. They provide important physical insights on the role of non-specific interactions in DNA loop formation.

\subsection{No desorption limit, $k_{\text{off}}(n) \rightarrow 0$}

If the DNA looped states are energetically strongly favorable ($G(n)<0$ and $|G(n)| \ge 1$ $k_{\text{B}}T$), then the protein will bind to DNA and it will not dissociate until the target sit is found. It can be realized, for example, if the protein-DNA non-specific interactions are very strong and attractive ($\epsilon <0$ and $|\epsilon| \gg 1$ $k_{\text{B}}T)$. This corresponds to $k_{\text{off}}(n) \rightarrow 0$ case,  which we call a no desorption limit.  Then  Eq. \ref{eq-M5} can be simplified into
\begin{equation}
    \frac{\partial F(n,t)}{\partial t}=-(\mu_n+\omega_n) F(n,t) +\mu_nF(n+1,t)+\omega_nF(n-1,t).
\label{eq-M7}
\end{equation}
In order to solve it together with Eq.\ref{eq-M6}, we apply the Laplace transformations, $\widetilde{F}(n,s)\equiv \int _{0}^{\infty} F(n,t)\exp(-st)dt $, where $s$ is the Laplace variable. Then Eq. (\ref{eq-M7})  transforms into
\begin{equation}
(s+\mu_n+\omega_n)\widetilde{F}(n,s)=\mu_n \widetilde{F}(n+1,s)+\omega_n \widetilde{F}(n-1,s)
\label{eq-M8}
\end{equation}
Correspondingly,  Eq.\ref{eq-M6} now can be written as
\begin{equation}
\left[s+ \sum_{n=1}^{L} k_{\text{on}}(n)\right] \widetilde{F}(0,s)=\sum_{n=1}^{L} k_{\text{on}}(n) \widetilde{F}(n,s)
\label{eq-M9}
\end{equation}
The most relevant quantity to describe the dynamics in the system is the mean search time $T_{n}$, which is defined as the average time to reach the target site $m$ when the initial binding site is at $n$,
\begin{equation}
 T(n)=\int_{0}^{\infty} t F(n,t)dt=-\frac{\partial \widetilde{F}(n,s)}{\partial s}\vert_{s=0}
 \label{eq-10}
\end{equation}
Correspondingly, the mean search time from the unbound state, which we label as a looping time, is given by
\begin{equation}
T=\int_{0}^{\infty} t F(0,t)dt=-\frac{\partial \widetilde{F}(0,s)}{\partial s}\vert_{s=0}.
\end{equation}
With the help of  Eq.\ref{eq-M9} it can be found that
\begin{equation}
 T= \frac{1}{k_{\text{on}}^{(S)}}+\sum_{n=1}^{L} \left(\frac{k_{\text{on}}(n)}{k_{\text{on}}^{(S)}}\right) T(n),
 \label{eq-11}
\end{equation}
where $k_{\text{on}}^{(S)} \equiv \sum_{n=1}^{L} k_{\text{on}}(n)$ is the total binding rate of the protein molecule to all DNA sites. The physical meaning of this result is the following. The total mean search time to reach the target from the unbounded state is a sum of two terms. The first terms describes the average time to bind to any site on DNA, while the second term is the average time to reach the target from the site $n$, $T(n)$ multiplied by the probability that the protein will associate to the site $n$ from the unbounded state. The coefficient $\frac{k_{\text{on}}(n)}{k_{\text{on}}^{(S)}}$ gives this probability.

To evaluate the looping time we need to calculate $T(n)$. This can be done in the following way. In this limit, the search process in the looped conformation can be viewed as a one-dimensional inhomogenous random walk, for which the first-passage times have been explicitly analyzed in terms of position-dependent hopping rates.\cite{kolomeisky2013}  We utilize these results for calculating $T(n)$  in Eq. (\ref{eq-11}).

\subsection{No sliding limit, $\mu_n=\omega_n \rightarrow 0$}

Another situation that can be solved analytically corresponds to the limiting case when the protein can form the transient DNA loops, but it cannot slide in the looped states. This can be associated with a very large free energy for being in the looped state ($G(n)>0$ and $|G(n)| \ge 1$ $k_{\text{B}}T$), and it might be realized for strong non-specific protein-DNA repulsive interactions,  ($\epsilon >0$ and $|\epsilon| \gg 1$ $k_{\text{B}}T)$. This corresponds to $\mu_n=\omega_n \rightarrow 0$, and we call this situation a no sliding limit.

Since this case has been fully analyzed previously,\cite{kolomeisky2016} here we briefly recapitulate the main results. Eq. \ref{eq-M5} in this limit is written as
\begin{equation}
    \frac{\partial F(n,t)}{\partial t}=-k_{\text{off}}(n) F(n,t)+k_{\text{off}}(n)F(0,t).
\end{equation}
In the Laplace domain, it transforms into
\begin{equation}
\left[s+k_{\text{off}}(n) \right] \widetilde{F}(n,s)=k_{\text{off}}(n) \widetilde{F}(0,s).
\end{equation}
With Eq. \ref{eq-M9} and the initial condition $\widetilde{F}(m,s)=1$, one can obtain the following expression,
\begin{equation}
\widetilde{F}(0,s)=\frac{k_{\text{on}}(m)}{s+f(s)}, 
\end{equation}
where the auxiliary function $f(s)$ is given by
\begin{equation}
 f(s)\equiv k_{\text{on}}(m)+\sum_{i\neq m}\frac{s k_{\text{on}}(i)}{s+k_{\text{off}}(i)}.
 \end{equation}
Then the mean search time $T$ can be easily computed, yielding
\begin{equation}
T=\frac{1+\sum_{i\neq m} \frac{k_{\text{on}}(i)}{k_{\text{off}}(i)}}{k_{\text{on}}(m)}.
\label{eq-15}
\end{equation}
This results underlines the fact that, on average, the protein should visit every site on DNA before the target can be found.

\subsection{No looping effect limit, $\mu_n=\omega_n \rightarrow const$}

There is one more limiting case that can be explicitly analyzed. If the free-energy associated with the formation of loops are relatively small, $ |G_{0}(n)| \leq k_{\text{B}}T$, then the search process is taking place in effectively flat free-energy profile. This was extensively investigated before for describing the single-site protein search.\cite{kolomeisky2013b,lange2015protein} 
Because in this case the transient formation of loops does not influence much the free energy of the system, we call it a no looping effect limit.

In this case, all transition rates become position independent, $k_{\text{on}}(n)=k_{\text{on}}$, $k_{\text{off}}(n)=k_{\text{off}}$ and $\mu_{n}=\omega_{n}=\mu$. Then it can be shown that the mean search time is given by
\begin{equation}\label{eq25}
T=\frac{1}{k_{\text{on}}}\frac{L}{S}+\frac{1}{k_{\text{off}}}\left(\frac{L}{S}-1\right),
\end{equation}
where a new parameter $S$ describes the number of sites visited during each binding event, and it depends on transition rates $k_{\text{off}}$ and $\mu$, see Refs.\cite{kolomeisky2013b} and \cite{lange2015protein} for more details. Eq. (\ref{eq25}) also has a clear physical meaning. There are $L/S$ protein bindings to DNA ($1/k_{\text{on}}$ is the time for each event), and there are $L/S-1$ unbindings ($1/k_{\text{off}}$ is the time for each event). The number of dissociations is less than the number of associations by one because the after last binding event the target will be found.

\section{Results}
\label{sec-results}
Now let us consider a general search problem for the two-site protein molecule already bound to DNA at the end of the chain to locate the second target sequence. We investigate it using Monte Carlo computer simulations with the Gillespie algorithm for various sets of parameters \cite{gillespie1977exact}. To describe the dynamics in the system, we introduce a new parameter $\lambda_0\equiv\sqrt{\mu_0/k_{\text{off}}}$, which we call a scanning length. It corresponds to a distance that the protein would explore while sliding along the DNA chain if diffusion rate at all sites will be the same and equal to $\mu_{0}$ and the dissociation rate will be the same and equal to $k_{\text{off}}$. The actual scanning length depends on the position of the binding, but it is always proportional to $\lambda_{0}$. Thus, the parameter $\lambda_{0}$ is a convenient measure of  non-specific protein-DNA interactions as well as the measure of the stability of the transient loop formation. The large the scanning length, the stronger is non-specific protein-DNA interaction and the longer the system is found in the looped conformation.

The results of Monte Carlo computer simulations, as well as analytical predictions in limiting cases, are shown in Fig. 3, where the looping time as a function of the scanning length is presented. Three dynamic regimes can be identified. If the scanning length is very small, $\lambda_{0} <1$, the protein occasionally binds to the DNA chain, but it cannot slide. This is a 3D search dynamic regime from the point of view of the protein molecule although it is always connected to DNA.  It was explicitly investigated before.\cite{kolomeisky2016} This also corresponds to the no sliding limit, considered above. Excellent agreement between analytical results and computer simulations in this regime shows that our theoretical arguments correctly capture the main physics in this regime. In the opposite limit of $\lambda_0 > L$ ($L$ is the length of the DNA chain), once the protein binds to the DNA, it remains on it until it reaches the target site. This is effectively a 1D dynamic process, and the search time $T$ is insensitive to the binding rate $k_{\text{on}}$ because the association occurs only once. Our analytical predictions also perfectly agree here with computer simulations. It is interesting to note that the dynamics in this regime might be faster or slower in comparison with $\lambda_{0}<1$ regime, depending on the association rates. If the binding rats are slow, then the 1D search is faster than 3D search because it needs only one binding event to reach DNA. However, when the binding rates are fast 3D search is more efficient since in the 1D regime the protein might be trapped by repeatedly moving over the sites that are far away from the target.

\begin{figure}
    \centering
    \includegraphics[width=0.9 \columnwidth]{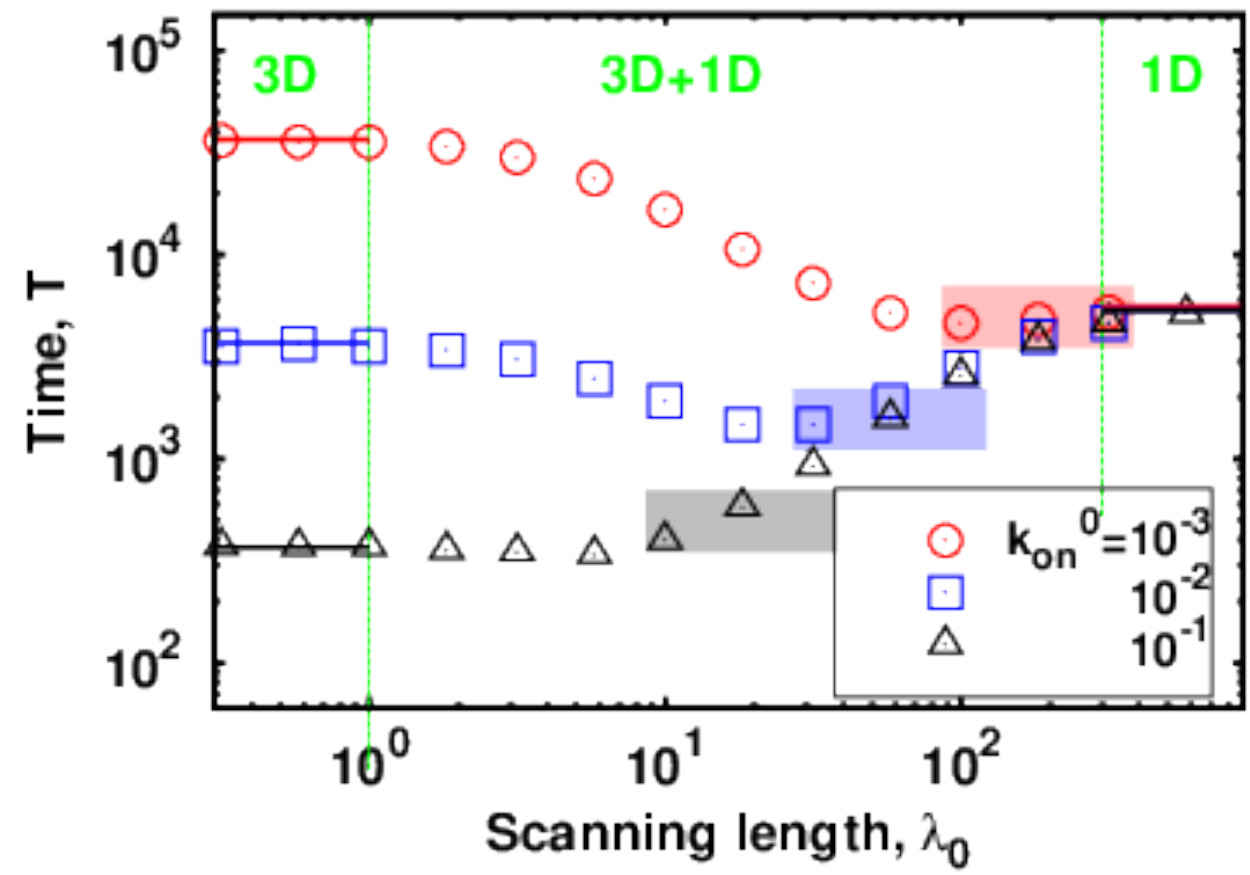}
\caption{Looping time $T$ as a function of the scanning length $\lambda_0$ for three different values of $k_{\text{on}}^{(0)}$. The target is located at the end of the chain $n=L$. Simulation data are shown in symbols and the solid lines are from theoretical predictions. The shadowed regions indicate the scanning lengths with biologically relevant attraction strengths $-5 k_{\text B}T \le \epsilon \le -2 k_{\text{B}}T$. For calculations we take $\mu_0=1$ and $L=100$.}
    \label{fig-3}
\end{figure}

The most interesting behavior is observed in the intermediate dynamic regime for $1 < \lambda_{0}<L$, which we label as 3D+1D search (see Fig. 3). In this case, the protein binds to DNA, slides some distance and dissociates, and then the cycle is repeated several times until the target is found. Computer simulations show that the search dynamic can be optimized in this dynamic phase. The minimum in the search time is observed for some intermediate scanning lengths. This physically corresponds to the situation when the protein is not trapped for a long time in sliding but can dissociate to start the search at a new location, but at the same time, it is not doing too many binding/unbinding events that might slow down the dynamics. It is interesting that the most realistic values of weak repulsions ($-5 k_{\text{B}}T<\epsilon<-2 k_{\text{B}}T$, shaded areas in Fig. 3) correspond to the region when the looping dynamic is the fastest. This suggests that nature might tune the non-specific protein-DNA interactions to speed up the formation of complex protein-DNA complexes.

Our theoretical approach allows us to quantify the role of transient loop formation in the overall search process. To do so, we view the free energy of the system as $G(n)=\epsilon+c G_{0}(n)$ and vary the parameter $0 \le c \le 1$. The case of $c=0$ describes a free diffusion in the flat free-energy profile that has been extensively employed for analysing single-site protein search dynamics.\cite{kolomeisky2013b,lange2015protein,shvets2015,shin2018b} For $c=1$, the DNA looping is fully taken into account as discussed above. The results of varying the elastic and entropic contributions associated with the transient loop formation are presented in Fig. 4. Generally, adding the loop formation slows down the overall search process. This can be understood by looking at the free-energy profile in Fig. 2. If the protein binds to the region closer to the minimum, then it takes much longer to explore the regions of DNA that correspond to higher values of the free energy. The protein also has a lower probability to attach to those regions. For the free diffusion (flat free-energy profile) this problem does not exist. However, there is a range of intermediate scanning lengths at which the dynamics with the loop formation is faster, see Fig. 4. It seems that for these parameters the dynamics with loop formation is faster because sliding is faster in the direction of the free-energy minimum, while the system is not trapped too long at the same region. Interestingly, this range of parameters also corresponds to the fastest search dynamics, emphasizing the potential importance of the loop formation due to non-specific interactions in cellular processes.

\begin{figure}
\centering
\includegraphics[width=0.9 \columnwidth]{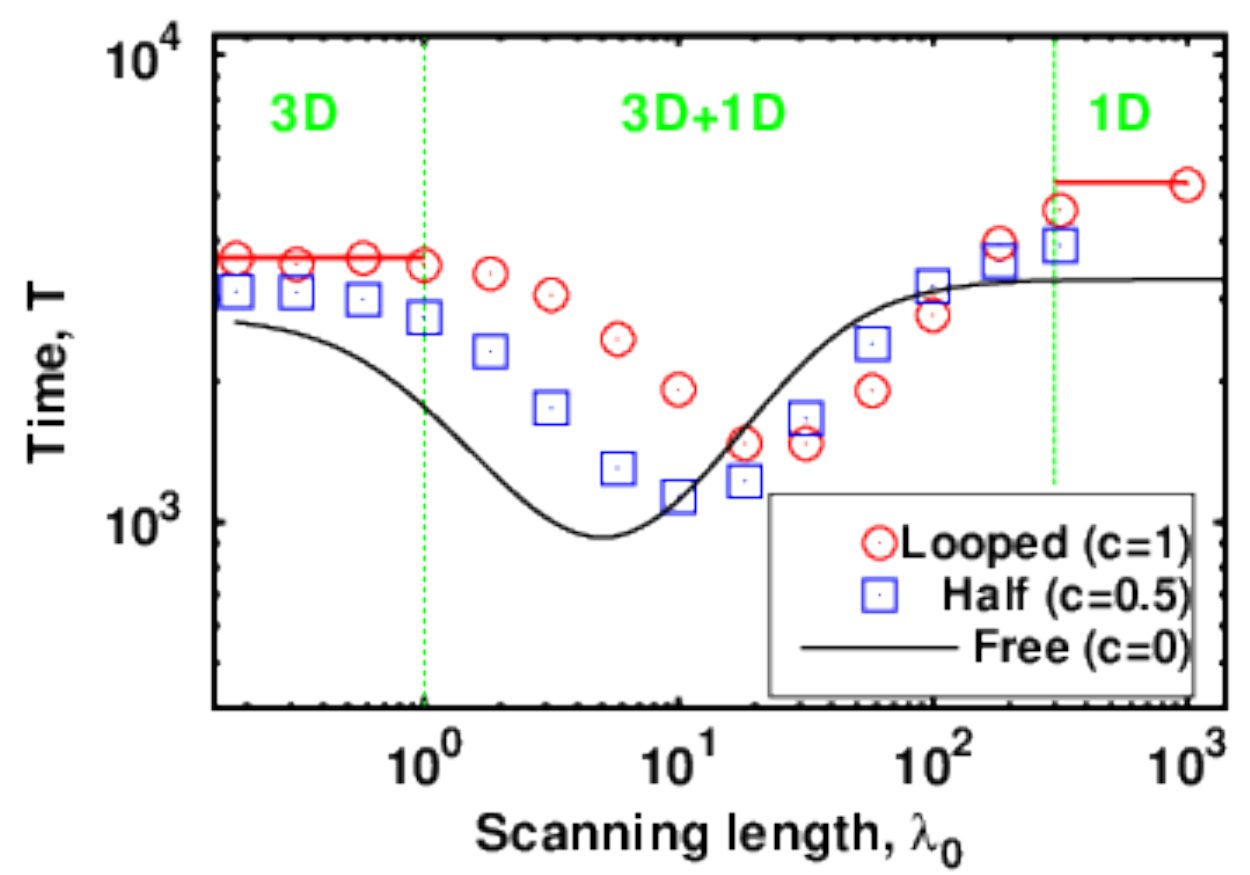}
\caption{Looping time $T$ as a function of the scanning length $\lambda_0$ by varying the contribution of looping in the free-energy profile, $G(n)=\epsilon+c G_{0}(n)$. Three cases for the parameter $c$ are considered: $c=0$ corresponds to the free diffusion (black line), $c=1$ corresponds to the search time with full contribution to the  free energy due to the loop formation (red symbols), and $c=0.5$ is an intermediate case (blue symbols). The target is located at the end of the chain $n=L$. For calculations we use $\mu_0=1$, $L=100$, and $k_{\text{on}}^{(0)}=0.01$.}
\label{fig-4-new}
\end{figure}

Because the free-energy profile generally is strongly position-dependent (see Fig. 2), it is reasonable to expect that the search dynamics will be sensitive to the location of the target. We investigated this effect, and the results are presented in Fig. 5 for different scanning lengths. As expected, the looping times depend on the target position $m$, however, this dependence is also determined by the nature of the dynamic regime. For small scanning lengths ($\lambda_{0}<1$, 3D search regime) the protein does not slide along the DNA chain and the probability of reaching the specific site on DNA is fully determined by the free-energy profile as given by Eq. (\ref{eq6}). The sites that are closer to the free-energy minimum are more probable to be explored first. For this reason, the dependence of the search time in 3D dynamic regime follows almost exactly the free-energy profile in Fig. 2. A different behavior is observed for large scanning lengths ($\lambda_{0}\geq L$, 1D search regime) when the protein associates only once with the DNA chain. In this case, the target can be achieved mainly via 1D diffusion. Then the average distance between the target and the location where the protein binds first to DNA determines the overall search time.  For this reason, the minimum search time is closer to $m=L/2$ position due to symmetry. For the intermediate 3D+1D dynamic regime, the overall search is faster, the scanning length showing the minimum looping time lies between two limiting cases, and the dependence on $m$ is weaker.

\begin{figure}
\centering
\includegraphics[width=0.9 \columnwidth]{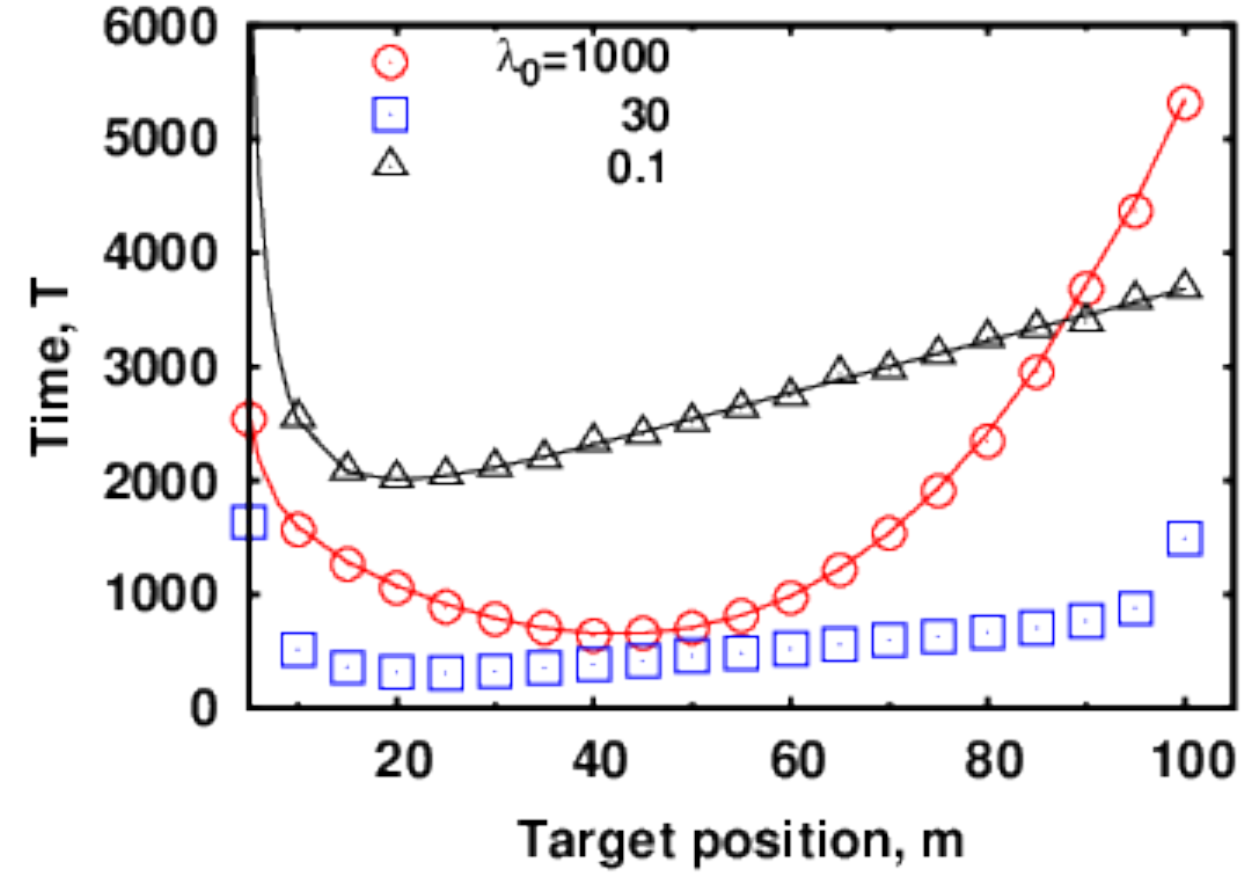}
\caption{Looping time $T$ as a function of the target position $m$ for different values of the scanning length $\lambda_0$. Simulation data are shown as symbols and theoretical predictions are shown as lines. For calculations we use $k_{\text{on}}^{(0)}=0.01$, $\mu_0=1$, and $L=100$.}
\label{fig-4}
\end{figure}

In our system, the process is taking place via the formation of transient polymer loops. But it is easier to form the loop for longer DNA segments than for the shorter chains. These arguments suggest that the DNA length $L$ might also be an important factor in the overall search process. We tested this idea, and the results are presented in Fig. 6. Here we show the looping time $T$ for three different values of $\lambda_0$. For all three cases, the looping time $T$ showed a minimum when the chain length corresponds to the loop size of the minimum in the free energy profile. The analytical theory for the 3D search (black line) matches excellently with the simulation data. The theory of 1D search (red line)  is also in a good agreement with Monte Carlo simulations. The presented results clearly show that the looping dynamics can be optimized by varying the DNA chain length.

\begin{figure}
\centering
\includegraphics[width=0.9 \columnwidth]{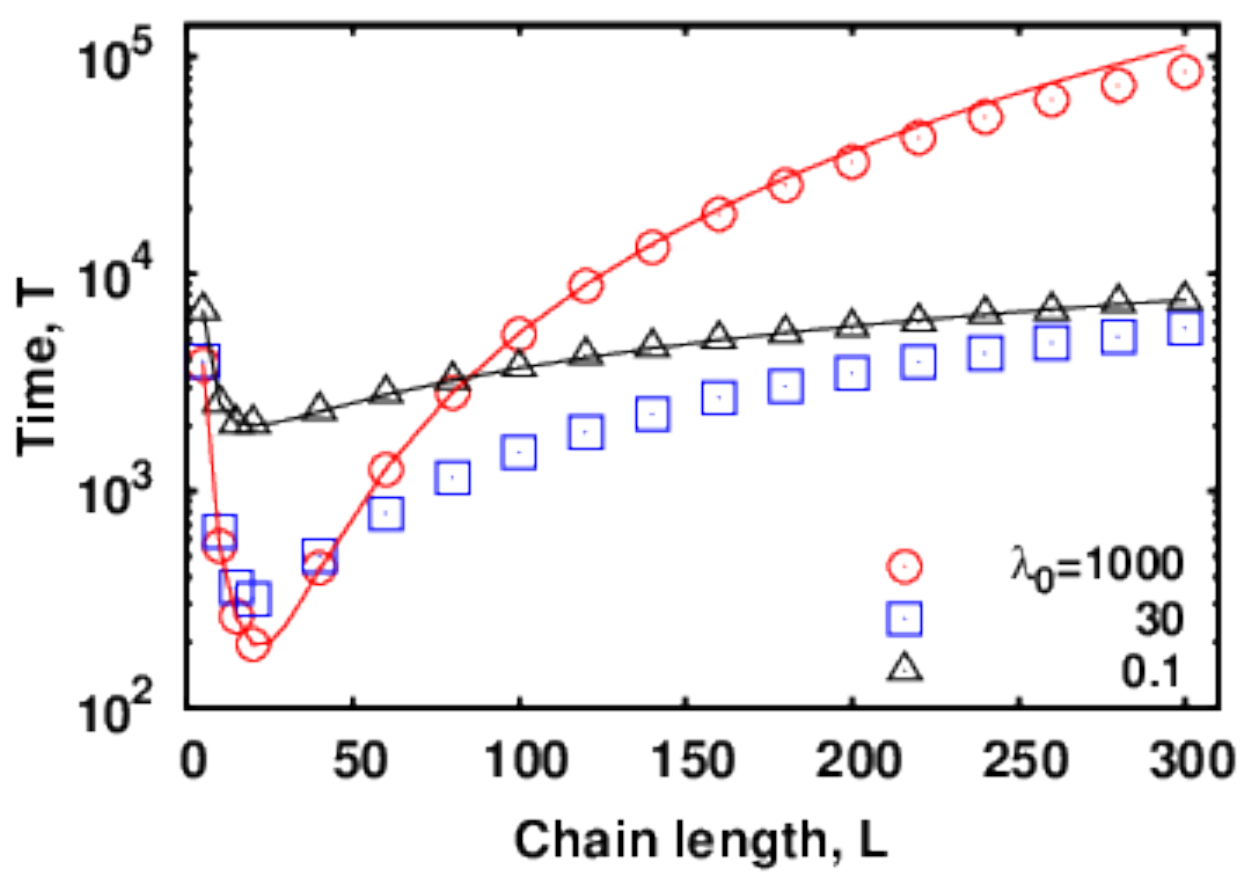}
\caption{Looping time $T$ as a function of the chain length $L$ for different values of $\lambda_0$. Simulation data are shown in symbols, and theoretical predictions are shown in solid lines. Here we take the parameters $k_{\text{on}}^{(0)}=0.01$, $\mu_0=1$, and the target is located at the end of the chain $m=L$.}
\label{fig-5}
\end{figure}

\section{Summary and conclusions}
\label{sec-conclusion}
We presented a theoretical analysis of the formation of a protein-DNA complex with a loop using analytical calculations and Monte Carlo computer simulations. We specifically considered two-site proteins that are already bound to DNA at one site that are searching for the second target site. A discrete-state stochastic model that takes into account the free-energy cost of the transient loop formation is utilized in our analysis. It is found that the non-specific protein-DNA interactions strongly influence the loop formation in the final complex. Three different dynamic regimes are identified depending on the relative sliding lengths and the size of the DNA chain. When the protein cannot slide along the DNA, the search is effectively three-dimensional with the formation and breaking of transient loops at each site. This corresponds to weak protein-DNA non-specific interactions. In the opposite limit of very strong non-specific interactions, after the first association to the DNA chain the protein slides continuously until the target is found. This is effectively a one-dimensional search. For the intermediate range of protein-DNA interactions, the slidings alternate with breaking and making transient polymer loops. It is found that the dynamics can be optimized (fastest) in this 3D+1D search regime. Our analysis shows the importance of the transient loop formation, and there is a range of parameters when it can even show faster dynamics in comparison with the case without loop formation. We also found that due to the free-energy changes associated with the formation of transient loops at different sites, the location of the target sequence affects the dynamics. In addition, the length of the DNA segment is another important factor in the formation of protein-DNA complexes due to different free-energy cost of making loops of different sizes. All these observations clearly show that the non-specific protein-DNA interactions are important in the formation of protein-DNA complexes with topological features such as loops.

Our theoretical approach is able to describe the main features of the non-specific interaction assisted DNA looping by multi-site proteins. However, it is worthwhile to discuss its limitations. One of the most important assumptions in our theoretical model is that the polymer relaxation is taking place much faster than other transition in the system. This assumption is supported by single-molecule measurements which show that protein diffusion on DNA is much slower than the free molecular motion in the bulk solution.\cite{elf2007} However, we did not take into account the sequence specificity of the DNA segments, while theoretical calculations indicate that this  might strongly affect the search dynamics.\cite{shvets2015} In addition, our theoretical model neglects protein and DNA conformational fluctuations that might complicate the process. Furthermore, real cellular systems are very crowded, and the presence of other molecules bound to DNA could prevent the search dynamics. We do not take this into account in our computations. Despite these limitations, it is reasonable to say that our theoretical method provides a consistent physical picture of the DNA loop formation with the help of non-specific protein-DNA interactions. The main advantage of our approach is quantitative predictions that can be tested in experiments. Therefore, it will be important to validate our results using various experimental techniques.

\section*{Acknowledgements}
This work was supported by the Welch Foundation (C-1559), by the NSF (CHE-1664218), and by the Center for Theoretical Biological Physics sponsored by the NSF (PHY-1427654).

%\balance

%%%REFERENCES%%%
\bibliography{shin-2019} %You need to replace "rsc" on this line with the name of your .bib file

\end{document}